\documentclass[twocolumn]{aastex631}

\usepackage{subcaption}
\usepackage{soul, xcolor}

\usepackage{mathtools,amssymb,amsmath}
\begin{document}

\setstcolor{red}

\title{The link between Athor and EL meteorites does not constrain the timing of the giant planet instability}

\author[0000-0003-1878-0634]{André Izidoro}
\affiliation{Department of Earth, Environmental and Planetary Sciences, 6100 MS 126, Rice University, Houston, TX 77005, USA}

\author[0000-0001-6730-7857]{Rogerio Deienno}
\affiliation{Southwest Research Institute, 1050 Walnut St. Suite 300, Boulder, 80302, CO, USA}

\author[0000-0001-8974-0758]{Sean N. Raymond}
\affiliation{Laboratoire d'Astrophysique de Bordeaux, Univ. Bordeaux, CNRS, B18N, all{\'e}e Geoffroy Saint-Hilaire, Pessac, 33615, France}

\author[0000-0001-8933-6878]{Matthew S. Clement}
\affiliation{Johns Hopkins APL, 11100 Johns Hopkins Rd, Laurel, 20723, MD, USA}      

%\collaboration{20}{(AAS Journals Data Editors)}



\begin{abstract}

The asteroid Athor, residing today in the inner main asteroid belt, has been recently associated as the source of EL enstatite meteorites to Earth. It has been argued that Athor formed in the terrestrial region -- as indicated by similarity in isotopic compositions between Earth and EL meteorites -- and was implanted in the belt $\gtrsim$60 Myr after the formation of the solar system. A recently published study  modelling Athor's implantation in the belt \citep{avdellidouetal24} further concluded, using an idealized set of numerical simulations, that Athor cannot have been scattered from the terrestrial region and implanted at its current location unless the giant planet dynamical instability occurred {\em after} Athor's implantation ($\gtrsim$60~Myr).  In this work, we revisit this problem with a comprehensive suite of dynamical simulations of the implantation of asteroids into the belt during the terrestrial planet accretion.  We find that Athor-like objects can in fact be implanted into the belt long after the giant planets' dynamical instability.  The probability of implanting Athor analogs when the instability occurs at $\lesssim15$~Myr is at most a factor of $\sim$2 lower than that of an instability occurring at $\sim100$~Myr after the solar system formation. Moreover, Athor's implantation can occur up to $\gtrsim$100 Myr after the giant planet instability. We conclude that Athor's link to EL meteorites does not constrain the timing of the solar system's dynamical instability.

\end{abstract}



\section{Introduction} \label{sec:intro}

It is widely accepted that the  the solar system's giant planets formed in a more compact orbital configuration and evolved to their current orbits via a dynamical planetary instability \citep{gomesetal05,levisonetal11,nesvornymorbidelli12,deiennoetal17}. The timing of this envisioned event, however, remains difficult to constrain~\citep{boehnk16,zellner17}. While it was originally suggested to have happened relatively late, about 500-800~Myr after the formation of the solar system~\citep[see][ and references there in]{gomesetal05,tsiganisetal05}, recently derived dynamical and cosmochemical constraints indicate that the instability happened within the first $\sim$100~Myr of the solar system history  \citep{morbidellietal18,nesvornyetal18,mojzsis19,ribeiroetal20}. The exact timing of the instability has dramatic implications for the early evolution of the inner solar system. If it happened earlier in this window \citep[potentiall as early as just after dispersal of the sun's gaseous disk;][]{liuetal22} it would have had a strong impact on the middle phases of terrestrial planet growth \citep{kaibchambers16,clementetal18}.  On the other hand, if it occurred later in this window it might have coincided with the timing of the Moon forming event \cite[e.g.][]{yinetal02,kleinetal09,desouza21}.

The timing of the giant planet instability also has important implications for the early evolution of the asteroid belt, Kuiper Belt, and other small body reservoirs~\citep[e.g.][]{nesvornyetal2021}.  Thus, strongly constraining the instability's timing has the potential to confirm or refute a number of different models and theories. It remains a central challenge to advance our understanding of solar system formation and evolution.

%would have strongly affected the formation of Mars (Kaib and Chambers 2016, Clement et al 2018, Deienno et al. 2018, Liu et al. 2022). 

%evolution of the Kuiper Belt and other small body reservoirs (Nesvorny 2015,2021).  Thus, additional constraints on the instability's timing have the potential conform or refute a number of different models.

%

%It may have happened during (or shortly after) the sun's natal disk dispersal \citep{kaibchambers16,clementetal18,liuetal22} or relatively later, when the terrestrial planets are already fully formed \cite[e.g.][]{yinetal02,kleinetal09}.

%and a subject of great scientific interest, given its broad implications for our  understanding of solar system formation and evolution.

%This has profound implications as it would exclude the possibility that the giant planet instability happened at the time of the gas disk dispersal~\citep{liuetal22}, for instance.

It was recently proposed  that the inner main belt asteroid Athor ($a\approx2.38$; $e\approx0.14$; $i\approx9$~deg) constrains the instability's timing to  $\gtrsim$60 Myr after the formation of the solar system \citep{avdellidouetal24}. Athor is a member of an asteroid family. Athor's family has recently been associated via spectroscopic observations as the source of rare EL type enstatite meteorites \citep{avdellidouetal22}.  Asteroid family identification and reconstruction techniques were used to suggest that Athor's parent body was originally about 64~km in diameter (Athor is today about 44~km) and that an impact event within the inner main belt about $3_{-0.4}^{+0.5}$~Gyr ago \citep{delboetal19} created its current asteroid family. Thermal evolution models and thermo-chronometric data of EL meteorites suggest that EL meteorites are fragments of an even larger planetesimal, with diameter of about 240-420~Km. In order to reconcile the proposed link between EL meteorites and Athor with these thermo-chronometer models, \cite{avdellidouetal22} suggested that Athor's parent body was a fragment of this same large planetesimal. The study further suggested that the break-up of the initial D$\sim$240-420~Km planetesimal that produced Athor's parent body occurred no earlier that $>$60~Myr after the condensation of calcium-aluminum- rich inclusions \citep[CAIs;][]{avdellidouetal22,trieloffetal22}. An earlier fragmentation would not be consistent with thermochronological data of EL meteorites \citep{trieloffetal22}.

If Athor's parent body indeed formed from the break-up of a D$\sim$240-420~Km object, it is highly unlikely that the formation event occurred within the modern main belt region.  That would result in an X-complex asteroid family population much larger than the current one, or produce relics of a ``lost'' asteroid family,  which are features not observed today \citep{avdellidouetal22}. Therefore, the most likely scenario, according to ~\cite{avdellidouetal22}, is that Athor's parent body formed when a large planetesimal (D$\sim$240-420~Km) was fragmented via an impact interior to the asteroid belt (e.g. $<$1.8~au) no earlier than $>$60~Myr after the formation of the solar system, and was subsequently implanted into the belt.  Athor itself is the largest surviving remnant of a second collision that occurred about a $\sim$Gyr after Athor's parent body's implantation into the belt \citep{delboetal19}. This secondary event formed the Athor family and made possible  the delivery the EL-type meteorites to Earth \citep{avdellidouetal22,trieloffetal22}.

As discussed in \cite{avdellidouetal22}, Athor's implantation from the terrestrial region seems to be further supported by the strong isotopic similarities bewteen EL meteorites and Earth ~\citep{javoyetal10,dauphas17}. The implantation of Athor from the terrestrial planet formation region into the asteroid belt is also broadly aligned with new solar system formation models suggesting that terrestrial planets formed from a narrow ring of planetesimals~\citep{hansen09,izidoroetal22,morbidellietal22}, and the asteroid belt may have been born `empty' -- that is, devoid of planetesimals \citep{raymondizidoro17b,izidoroetal22,izidoroetal24}. Athor is potentially the best evidence that these models may be plausible.

%Asteroids from the terrestrial planet region can be implanted in the belt during the accretion of terrestrial planets if the terrestrial planets formed from a narrow ring of planetesimal ~\citep{raymondizidoro17b,izidoroetal24}.

Building on the arguments and results of \cite{avdellidouetal22}, \cite{avdellidouetal24} suggested that Athor can not have been implanted from the terrestrial region into the belt -- near to its current location -- if the giant planet dynamical instability occurred earlier than $\sim60$~Myr after the formation of the solar system. To rule out a dynamical instability at $\lesssim60$~Myr, \cite{avdellidouetal24} performed a set of numerical simulations modelling the implantation of planetesimals in the asteroid belt considering different idealized dynamical scenarios. In their nominal simulations, they did not model the accretion of terrestrial planets as we do here, but instead assumed that the terrestrial planets were fully formed and interacting with a leftover population of planetesimals crossing the asteroid belt. A set of their simulations also included the effects of a leftover embryo on a moderately eccentric orbit crossing the asteroid belt, which they refer to as Theia. From these simulations, \cite{avdellidouetal24} concluded that the only plausible way to implant Athor into the belt at such a late epoch (between $\sim$60 and $\sim$100  Myr after the formation of the solar system) is if the planetary dynamical instability occurred after Athor's implantation (i.e., $>$60 Myr after CAIs formation). We will further discuss the differences between our simulations and theirs later in the paper.

%We acknowledge that the link between ELs and Athor and its family members found by \cite{avdellidouetal22} is remarkable but it remains unclear whether this link is unique. Athor's family is several Gyr old and most likely lost a large fraction of its members by collisional evolution and dynamical diffusion, potentially even the large parent bodies or largest remnants \citep{delboetal17,delboetal19}. Giving the large uncertainties involved in associated methods and family identification techniques, we argue that some caution is required when linking asteroid, asteroid families, meteorites and thermochronological data. 

%

%As we have demonstrated in this work, the later the giant planet instability occurs, the larger is the planetesimal implantation efficiency into the belt.

%Finally, it is dramatically dangerous to attempt to rule out  a specific dynamical origin for a single asteroid invoke  fairly idealized dynamical scenarios to rule out

In this paper, demonstrate that a much broader range of instability times are consistent with Athor's implantation into the belt at $\gtrsim60$~Myr. Plausible scenarios include dynamical instabilities that took place during the first $\sim$15 Myr of the solar system evolution. We show that the simulations of  \cite{avdellidouetal24} failed to capture key dynamical processes that commonly take place during the accretion of terrestrial planets and that influence the capture of planetesimals in the belt. Furthermore, we discuss the challenges involved in using metrics derived from a single small body to constrain key aspects of the solar system's early evolution such as the timing of the giant planet instability

%We demonstrate that any individual asteroid alone can not be used to trace the giant planet instability timing.

Our paper is structured as follows. In Section \ref{sec:methods} we describe our methods and numerical simulations. In Section \ref{sec:results} we present our results, and in Section \ref{sec:conclusion} we present our final conclusions.

%Any individual asteroid alone can not be used to trace the giant planet instability timing.

\section{Methods}\label{sec:methods}

Our simulations model the accretion of terrestrial planets within a narrow ring of planetesimals.  These simulations were first presented and analyzed in \cite{izidoroetal24}; here we explore the relevance in the context of Athor's implantation specifically. They were conducted in two steps. First we modeled the growth from planetesimals to planetary embryos \citep{kokuboida00} using the LIPAD code~\citep{levisonetal12}. Each ring was assumed to have a specific radial surface density profile, proportional to $\Sigma(r)=r^{-x}$, and different simulations considered different values of x (0, 1, and 5.5). We also performed simulations with an ``upside-down U-shape'' ring profile, represented by $\Sigma(r)=(-200(r/{\rm au}-1)^2+24){~\rm g/cm^2}$ \citep{izidoroetal22}. Our LIPAD simulations started with 3000 planetesimals \citep{izidoroetal22}, which collisionally evolved following the LIPAD prescription. We neglected the effects of gas-driven planet migration in all simulations, but we accounted for the effects of aerodynamic gas drag~\citep{adachietal76}, as well as inclination and eccentricity gas-tidal damping~\citep{cresswellnelson06}. The gas disk lifetime was set to 5~Myr \citep[e.g.][]{willianscieza11}, and it dissipated following an exponential decay with e-fold timescale of 2~Myr. Our simulations were integrated using a time-step of 4 days.

In the second step, we used the outcome of our LIPAD simulations as inputs to simulate the growth from planetary embryos to final planets \citep{chamberswetherill98}. For step-2, we used the Mercury integration package \citep{chambers99} and assumed that the sun's natal disk was already dissipated. Objects more massive than the Moon were treated as self-interacting particles. Lower mass objects did not self-interact, but gravitationally interact with the sun and other massive objects. Our step-2 simulations started with approximately 300 to 500 planetesimals and 20 to 30 planetary embryos, depending on the planetesimal ring surface density profile~\citep[see][for details]{izidoroetal24}.

The timing of the instability was treated as a free parameter. We performed simulations with instability times ($t_{\rm inst}$) of 0, 5, 10, 50 and 100 Myr relative to the time of gas disk dispersal (start of step-2). Assuming that the sun's natal disk dissipated 5 Myr after the condensation of CAIs, our instability times correspond to  5, 10, 15, 55 and, 105 Myr after the solar system's formation, i.e, the time of CAIs condensation \cite{amelinetal10}.  Unlike our previous work \citep{izidoroetal24}, in this paper we exclusively express time relative to the birth of the solar system formation (CAI formation). We performed 50 simulations, starting with slightly different distributions of planetary embryos and planetesimals, for each combination of instability time and ring surface density profile.

We used two different approaches to simulate the dynamical instability of the giant planets.  In the first one 
we model the instability using numerical interpolation of a giant planet dynamical evolution which has been shown to be broadly consistent with solar system constraints \citep{deiennoetal18}. We refer to these simulations as ``Interpolated instability''. In the second case, we model the giant planet instability assuming the giant planets instantaneously ``jump''  from pre-instability orbits to their current ones.  We refer to these simulations as ``Instantaneous Instability''. This latter approach is common in the literature \citep[e.g.][]{bottkeetal12,brasseretal13,deiennoetal16}, and allows  us to circumvent potential specific effects associated with  weakly constrained variables when modelling the giant planet dynamical instability. For instance, the initial resonant configuration of the giant planets, the total mass assumed of the primordial planetesimal disk beyond Neptune, the extension of the planetesimal disk, etc. are weakly constrained parameters in giant planet instability instability models \citep[e.g.][]{nesvorny11}. Different assumptions for these parameters   may  lead to very different dynamical evolution pathways for the giant planets themselves. The main advantage of invoking the ``Instantaneous instability'' approach is that it allows us to mitigate any specific dynamical effect coming from the assumed interpolated instability itself.

Our ``Interpolated instability'' simulations were numerically integrated until the end of the instability phase only ($t_{\rm inst}$+0.8~Myr). Our ``Instantaneous instability'' simulations were integrated up to 200 Myr following gas disk dispersal. We do not extend the integration time of our ``Interpolated instability'' simulations up to 200 Myr in order save cpu-time (for details see \cite{izidoroetal24}).

Before the dynamical instability, Jupiter and Saturn were initially placed in the 3:2 mean motion resonance \citep[see also \citet{deiennoetal17}]{massetsnellgrove01,morbidellicrida07} with $a_{\rm J}=5.4$~au, $a_{\rm S}=7.3$~au, $e_{\rm J}=e_{\rm S}=0$, and $i_{\rm J}=0$, and $i_{\rm J}=0.5$~deg. After the instability, their orbits are represented by $a_{\rm J}=5.25$~au, $a_{\rm S}=9.54$~au, $e_{\rm J}=0.048$, $e_{\rm S}=0.056$, and $i_{\rm J}=0$, and $i_{\rm J}=1.5$~deg.

\section{Results}\label{sec:results}

We first present results from the interpolated instability simulations (Section 3.1), followed by results from the instantaneous instability simulations (Section 3.2).  Given the uncertainties in the exact evolution of the giant planets during the instabilities, we consider that each of these scenarios carries equal weight.  

\subsection{Interpolated Instability}

%The analysis of our results comes in two flavors. 

In this section, we first analyzed the implantation of planetesimals into the Athor region (hereafter defined as $2.3<a<2.5$~au; $e\leq0.15$; $i\leq15$~deg) in  simulations with interpolated instabilities. We remind the reader that we do not follow the full accretion phase of the terrestrial planets and long term evolution of the solar system in our interpolated simulations. These specific simulations stop shortly after the instability phase ($t_{\rm inst}+0.8$~Myr). At the end of these simulations, we look for Athor-like objects in the asteroid belt and constrain their implantation efficiency. Only in our subsequent section, dedicated to simulations modelling the instability as an instantaneous event, we do constrain the efficiency of implantation during the epoch of terrestrial planet formation \textit{after} the instability. We choose to present these results separately in order to improve our statistics and compare the results of these two different instability approaches. 

To start our analysis of the interpolate instabilities, we neglect the constraint suggested by \cite{avdellidouetal22} that the Athor must be implanted in the belt at least $\sim$60~Myr after the formation of the solar system.  We will return to this point later. Our goal at this point is simply to check whether  objects can be implanted in the Athor region in our various simulations with different instability times or not.

%5Myr - r0 (1)= 1 
%10Myr -r0 (1), r1(1) = 2 
%50Myr -r0 (4), r1(4), r55(1), ushape(1) = 10 
%100Myr -r0(2),r1(2),ushape(1) = 5

Altogether,  the initial number of planetesimals in our simulations at 5 Myr is $\sim$80~K planetesimals. We report probabilities and implantation efficiencies relative to this initial number of planetesimals.

\begin{figure}
\centering
\includegraphics[scale=0.45]{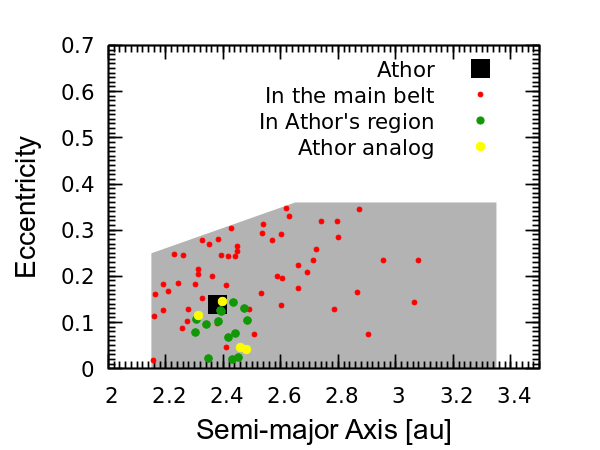}
\includegraphics[scale=0.45]{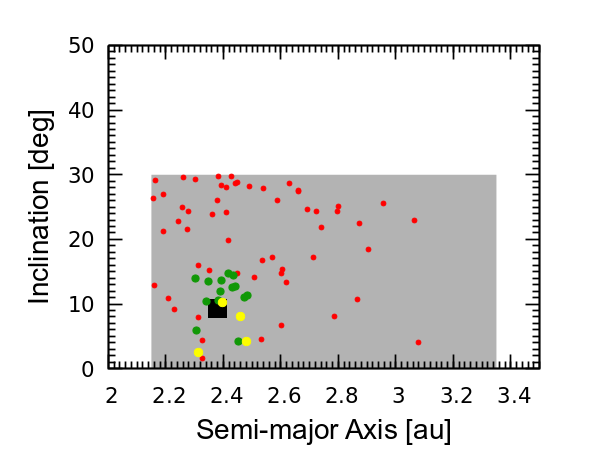}
\caption{Distribution of planetesimals implanted in the main belt and Athor's region from 17 selected simulations with interpolated instabilities. We plot together all  simulations that implanted at least one planetesimal in the Athor-region. Yellow dots show planetesimals that were implanted in the Athor region $>$60~Myr after the formation of the solar system and, therefore, are consistent with Athor (Athor analogues). Green dots show planetesimals implanted in the Athor region before 60~Myr. Red dots show asteroids implanted in the main belt. Athor's orbit is represented by the black square.}
    \label{fig:asteroid}
\end{figure}

Our interpolated simulations with $t_{\rm inst}=$ 5~Myr did not implant any Athor-like objects in the belt. Simulations with $t_{\rm inst}=$ 10~Myr, 15~Myr, 55~Myr and 105~Myr implanted 1, 2, 10, and 5 objects with Athor-like orbits, respectively. The number of Athor-like objects produced in simulations with different ring surface density profiles were: 8 planetesimals for $r^0$,  7 planetesimals for  $r^{-1}$,  2 planetesimals for ``U-shape'' ring profile , and 1 planetesimal for $r^{-5.5}$. These objects are shown in Figure \ref{fig:asteroid}. Considering all implanted planetesimals, in all our simulations with instability times occurring at anytime between 10 and 105~Myr, the average implantation probability was about 18/80,000~$\approx$2.2$\times10^{-4}$ .  The  average implantation probability for $t_{\rm inst}\lesssim$15~Myr is $\approx$3.7$\times10^{-5}$, and for  $t_{\rm inst}\gtrsim$55~Myr is $\approx$1.9$\times10^{-4}$.  

These probabilities do not account for the timing of implantation. In order to effectively determine whether our implanted planetesimals are truly consistent with Athor or not, we must further constrain our sample to those implanted in the belt late enough to be consistent with the thermo-chronometer modeling of \citet{trieloffetal22}.

%These implantation efficiencies are given at the end end of the instability phase, which lasts about 0.8 Myr. 

%We recall that we stop our interpolated simulations at the end of the instability phase. 
%In order to be able to compare these results with those of the next Section --  where we present the implantation efficiencies at the end of the accretion of terrestrial planets -- one should account for an additional depletion of a factor of $\sim$3 to $\sim$12, which is  expected to occur in the inner main belt ($a<2.5$~au) for $t_{\rm inst}=55$~Myr and 105~Myr~\citep[see Tables 2 and 3 of ][]{izidoroetal24}. Envisioning a subsequent  depletion of a factor of 3 of the Athor region during the accretion of terrestrial planets, these efficiencies become $\approx$1.2$\times10^{-5}$ for $t_{\rm inst}\lesssim$15~Myr, and $\approx6.3\times10^{-5}$ for  $t_{\rm inst}\gtrsim$55~Myr. Finally, 

We follow the arguments of \cite{avdellidouetal24} and assume that Athor-like objects should be implanted in the asteroid belt later than $\sim$60 Myr after CAIs, after the break-up of its larger parent body as suggested by thermo-chronometer models~\citep{trieloffetal22,avdellidouetal22}. By construction, our interpolated simulations with $t_{\rm inst}\lesssim$15~Myr cannot be used to test if they can implant Athor-like objects or not because they stop at about $t\lesssim$15+0.8~Myr. However, we can use our simulations with $t_{\rm inst}=55$~Myr and 105 Myr to look into this.

Our simulations with  $t_{\rm inst}=55$~Myr implanted 10 planetesimals in the Athor region, for an implantation efficiency of  $\approx$1.2$\times10^{-4}$. Nevertheless, a closer inspection shows that all ten planetesimals were already transiting across the belt region (on orbits where $1.8<a<3.5$~au, $i<30$~deg and $e\lesssim0.35$) before the instability time. These objects are shown in green in Figure \ref{fig:asteroid}. These planetesimals entered the belt region between 25 and 45 Myr, with an average entrance time of 40 Myr.  Consequently, although they were not effectively implanted until the instability finally occurred,  these planetesimals are unlikely to be consistent with Athor because their eventual break-up inside the belt (even if at $\gtrsim$ 60 My after CAIs)  would leave features which are not observed today~\citep{avdellidouetal22}. 

In simulations with $t_{\rm inst}=105$~Myr,  5 planetesimals were implanted in the Athor region. These planetesimals entered the asteroid belt region between $\sim$25 and $\sim$60 Myr, with an average entrance time in the asteroid belt of $\sim$53~Myr. In these simulations, 4 out of the 5 planetesimals reached the asteroid belt at about $\sim$60~Myr making them, at best,  marginally consistent with EL meteorites and Athor. These objects are shown in yellow in Figure \ref{fig:asteroid}. This yields an implantation efficiency in the Athor region of roughly $\approx$5$\times10^{-5}$. However, because no planetesimal was implanted  at $\gg60$ Myr in the Athor region in any of our simulations  with $t_{\rm inst}=105$~Myr, one can conclude from these simulations that the implantation of Athor at $\gg$60 Myr may be an event with a probability as low as $\lesssim$1.2$\times10^{-5}$. This  suggests that implanting Athor at $\sim$100 Myr in the framework of a late instability is in fact quite  challenging. 

It is important to keep in mind that these reported implantation probabilities might be misleading since they could be positively or negatively weighted by simulations that may or may not turn out to be good solar system analogs at the end. There are also many different ways to compute and report implantation probabilities. For instance, one may decide to account only for the total number of planetesimals in a specific  simulation (or set of simulations) with successful implantation. Our simulations typically start with 300 to 500 planetesimals. If a single planetesimal is implanted in a specific simulation, its implantation could be given as  1/500, which is a much higher value than our average probability, reported over our full sample of simulations (with 80,000 planetesimals). That being said, because we have only a single Athor asteroid, any non-zero probability is in fact quite  reassuring.

In the next section, we will present the implantation probabilities of selected solar system analogues produced in simulations with ``Instantaneous Instability''. The primary goal of the next section is to demonstrate that Athor-like objects can be implanted in the asteroid belt even if the giant planet instability occurred very early, for instance around the time gas disk dispersal \citep[e.g.][]{liuetal22}.  We will use these same simulations to compute the probability of Athor's late implantation for a range of different instability times.

%and also whether the system happens to experience the dynamical mechanism of a late embryo being kicked out of the Mars region.  The point being, whats more important is that it can happen for any of the instability times, and we only have 1 Athor so any non-zero probability is encouraging

%We conclude that this section by stating that the implantation efficiency of Athor-like objects at/after $\gtrsim$60 Myr in the context of a  relatively late instability ($t_{\rm inst}=$105~Myr) is $\approx$1.7$\times10^{-5}$.

%If we assume that all these planetesimals are consistent with Athor, we can conclude that the implantation efficiency of Athor-like objects at/after $\gg$55 Myr in the context of a  relatively late instability ($t_{\rm inst}=$105~Myr) is $\approx$5$\times10^{-5}$ ($\approx10^{-5}$ when accounting for a factor of 5 depletion during the accretion of the terrestrial planets), but if these objects are not consistent with Athor it could be as low as $\lesssim$1.2$\times10^{-5}$.

\subsection{Instantaneous Instability}

We now turn our attention to the simulations in which the giant planet instability was modeled as an instantaneous jump in their orbital elements, as described above. 
These simulations were numerically integrated for 200 Myr.

Altogether, out of 200 simulations (50 for each ring profile) starting with rings with surface density profiles proportional to  $r^0$, $r^{-1}$, $r^{-5.5}$, and ``U-shape'' and  $t_{\rm inst}=$ 5 Myr, 3  planetesimals were implanted in the Athor-region (one planetesimal per simulation). This corresponds to an implantation efficiency of $\approx3.7\times10^{-5}$ at 200 Myr.  These three simulations produced fairly good solar system analogues, with small Mars analogues and at least 3 terrestrial planets. Planetesimals were implanted in the Athor region after the instability took place ($t_{\rm inst}=$ 5 Myr). We have verified that the three implanted planetesimals  of these simulations typically entered the asteroid belt between about 28 and 40 Myr, with an average entrance time of 33 Myr after CAIs formation.

% We recall that our {\it interpolated} instability simulations with $t_{\rm inst}=$ 5 Myr (see previous section) did not show any planetesimals in the Athor region. We stop the interpolated instability simulations shortly after the instability phase. This indicates that, in our instantaneous instability simulations which were integrated from the instability time up to 200 Myr,

We extended the integration time of these simulations  with $t_{\rm inst}=$ 5 Myr to 1 Gyr, and only one of them produced a long-term stable implanted object. We will refer to this specific simulation in this section as ``Athor-1''. As no planetesimal was implanted in the Athor region after 60 Myr in any of these 200 simulations with, we can already expect that the probability of implanting Athor after 60 Myr is relatively low, and roughly of the order of $\lesssim1.2\times10^{-5}$. Later in this paper we will confirm that Athor's implantation probability after 60 Myr for an early instability scenario  (e.g. $t_{\rm inst}=$ 5 Myr) is about $\approx2\times10^{-5}$. 

With the goal of improving the statistics of our simulations when probing the long-term stability of Athor-like objects, we performed additional simulations in which we cloned selected simulations that produced good solar system analogues.  Mars-analogues are defined as planets with the following orbital and physical parameters: semi-major axes between $1.3<a<1.8$ au; orbital eccentricities lower than $\leq0.1$; orbital inclinations lower than $i\leq10$~deg, and masses between $0.05<M<0.3~M_{\oplus}$. Earth-analogues are defined as planets with masses higher than 0.5 $M_{\oplus}$,  orbital eccentricities lower than $\sim$0.1, and orbital inclinations lower than $\sim$5 degrees. We do not define Mercury and Venus analogues, but some of our selected simulations also produced reasonable analogues of these planets, with $a$, $e$, $i$, and $mass$  within a factor of $\sim$~two of the real planets.

We cloned our selected systems by changing planetesimals' semi-major axes by a tiny fraction ($\pm$0.001\%). This cloning process allows us to increase our simulation sample size and to save CPU time, instead of running new simulations from the beginning to increase our simulation number.

We start by cloning our ``Athor-1'' simulation. We focus on it exclusively in the subsequent paragraphs and analyses. We cloned this simulation 100 times at 70 Myr, when there were about 30 leftover planetesimals in the system and the terrestrial planets were not fully formed yet. We integrated each clone simulation up to 200 Myr. For the most successful cases (i.e. when an object is implanted in the Athor-region), we further extended the simulation to 1 Gyr. 

\begin{figure}[h]
\centering
\includegraphics[scale=0.56]{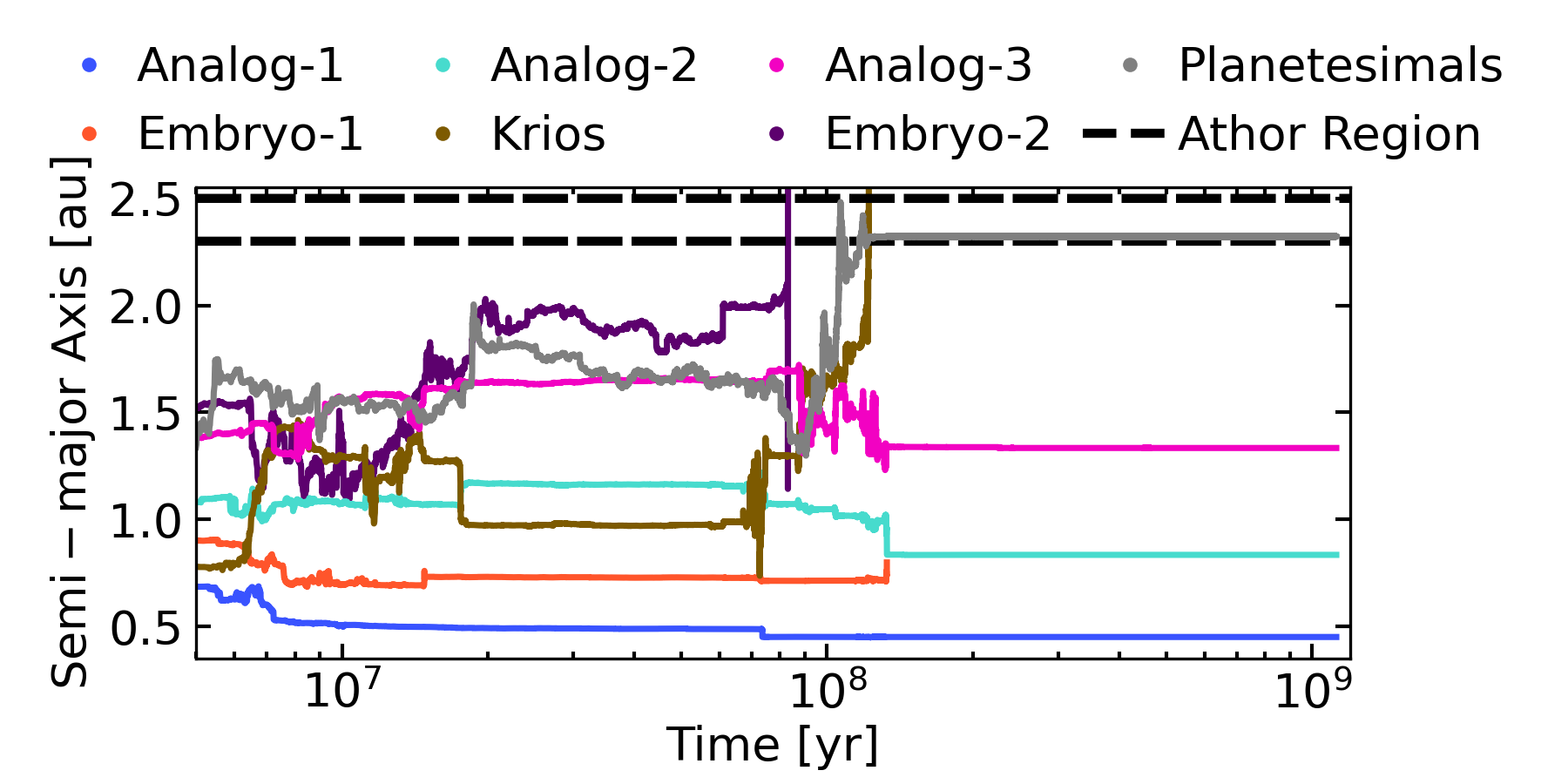}
\includegraphics[trim={0 0 0cm 1.2cm},clip,scale=0.56]{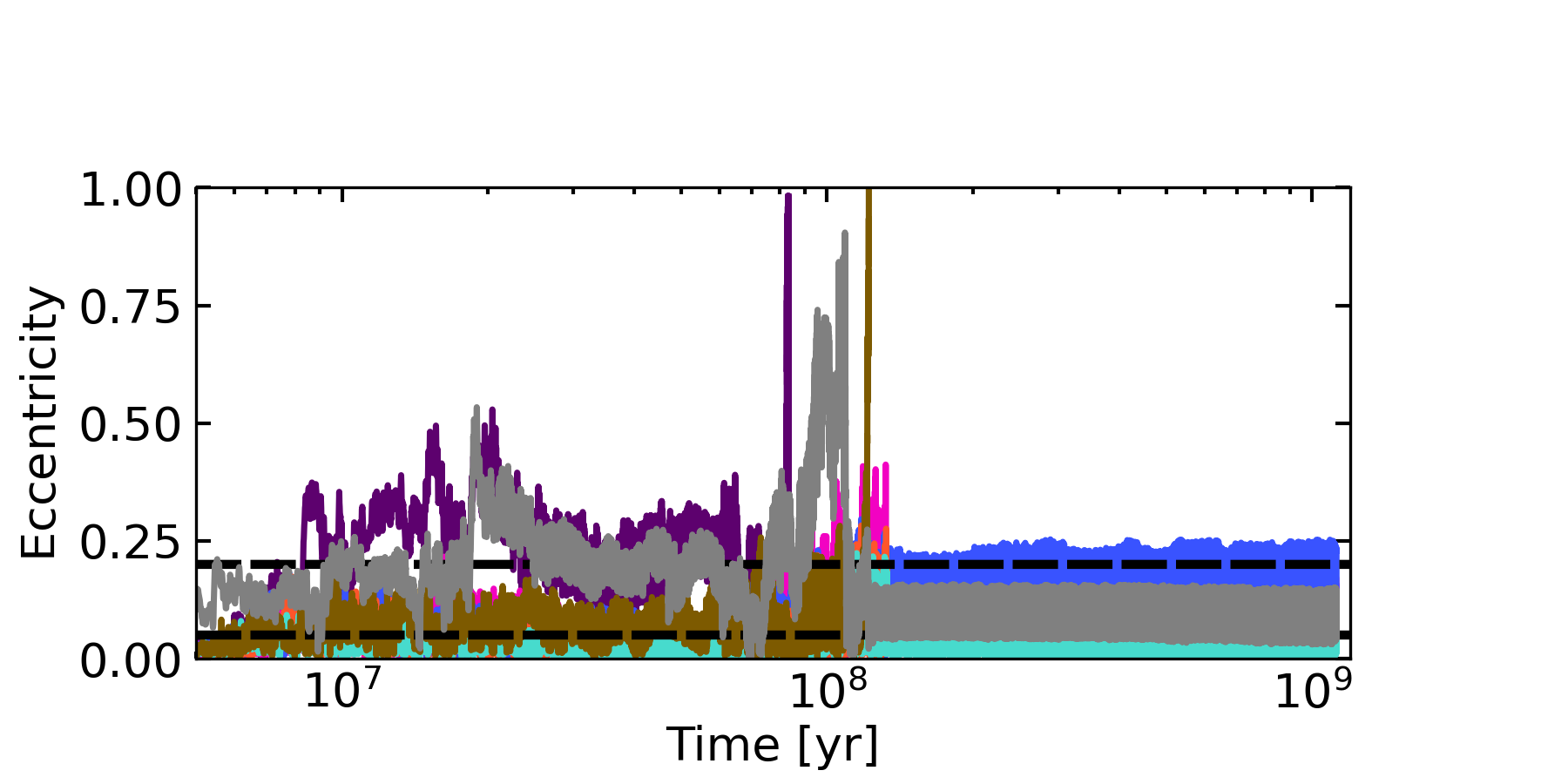}
\includegraphics[scale=0.56,trim={0 0 0cm 1.5cm},clip]{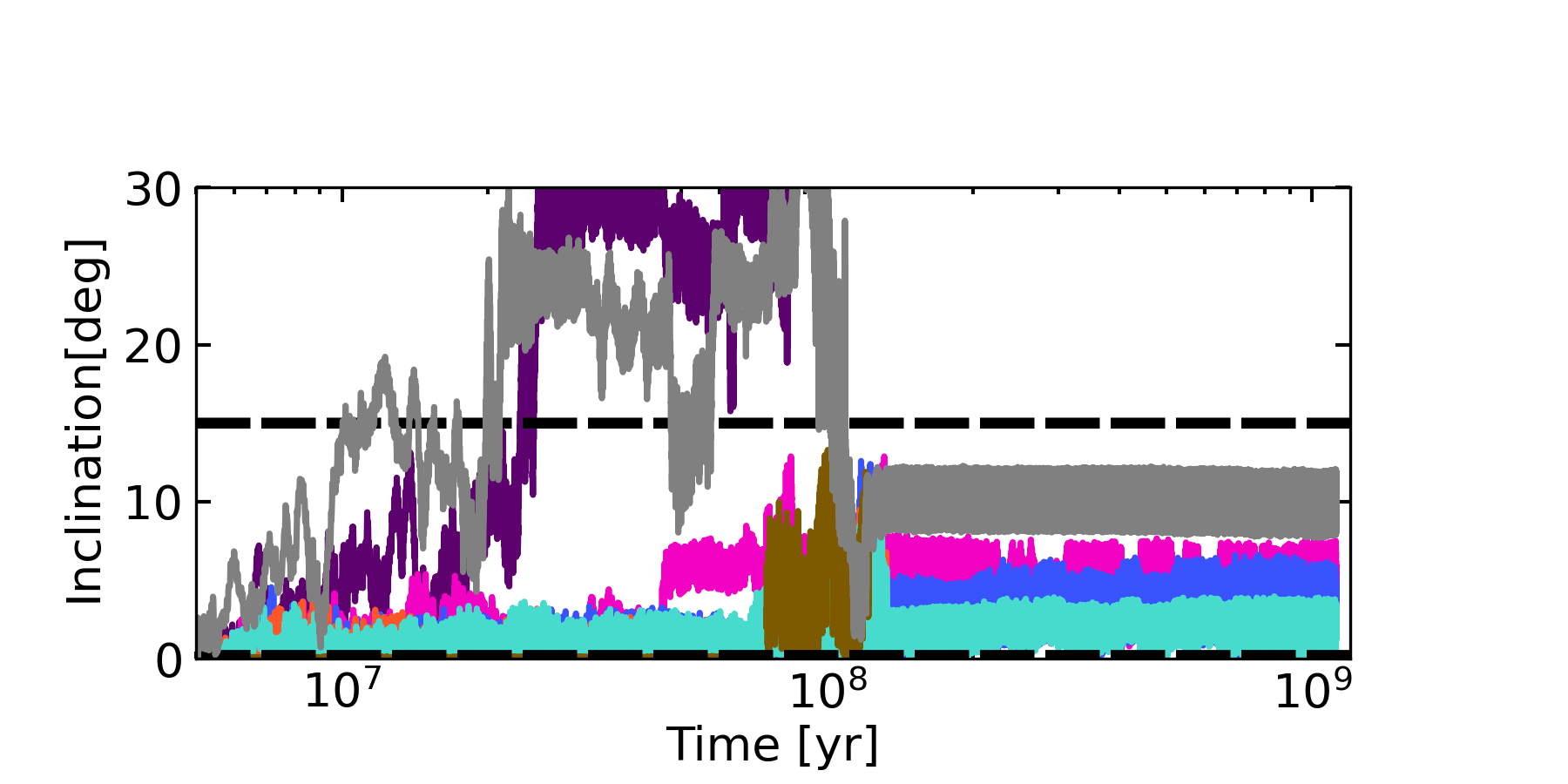}
\caption{Implantation of a planetesimal from the terrestrial region into the asteroid belt during the final accretion phase of terrestrial planets in our Athor-1 simulation. The planetesimal is implanted in the inner main belt, in the Athor-region. The x-axes shows time and the y-axes shows, from top-to-bottom, semi-major axis, orbital eccentricity, and inclination. Dashed lines  delimit the Athor region defined as $2.3<a<2.5$~au, $e<0.15$, and $i<15$~deg. This simulation corresponds to an example where the instability takes place at the time of the disk dispersal, i.e., $t_{\rm inst}=$5~Myr after the solar system formation. At 5~Myr, the grey-line planetesimal is at about 1.4~au and it finally enters the asteroid belt at $\sim$100 Myr ($a\gtrsim1.8$~au). Its effective implantation  occurs at $\sim$120~Myr, when an planetary embryos (which will be later refereed to as Krios) (light-brown) is ejected from the system.}
    \label{fig:analog1}
\end{figure}

From this set of 100 clone simulations, 4 simulations implanted an object in the Athor-region (one planetesimal per simulation). 3 out of the 4 implanted planetesimals were the same object  implanted in our original Athor-1 simulation. In all 3 cases, this object survived in the belt for 1 Gyr. However, one new planetesimal (which was ejected in the original  Athor-1 simulation) was implanted in the Athor region in one of the clone simulations. Figure \ref{fig:analog1} shows the implantation of this object and its long-term dynamical evolution for 1 Gyr. This clone simulation produced three terrestrial planets (Analogs 1 through 3). Note that the planetesimal implanted in the Athor-region, shown in grey, is at about 1.4~au  at 5 Myr and finally reaches the asteroid belt at about $\sim$100 Myr. This late implantation in the belt makes this object consistent with Athor~\citep{avdellidouetal24}, i.e., it lived in the terrestrial region ($a<$ 1.8 au) for $>$ 60 Myr before implantation.  We can also use this set of clone simulations to reassess the probability of implanting such an object at late time.

Envisioning that each of our clone-simulations would have started with about 500 planetesimals within 1.5~au (see Section \ref{sec:methods} and \cite{izidoroetal24}) as our original ``Athor-1'' simulation, the overall probability of implanting this object is $\approx$2$\times10^{-5}$ (1 object implanted/(100 clones$\times$500 planetesimals)). This represents the probability of implanting an Athor-like object at $\sim$100~Myr for an instability taking place at $\sim$5~Myr.

%These results combined with those of our interpolated instabilities suggest that the probability of implanting an object in Athor region before 60 Myr is about two to three times higher than between $\sim$60 and $\sim$100 Myr, for dynamical instabilities occurring at   $t_{\rm inst}=$ 5-15 Myr.

\begin{figure}[h]
\centering
\includegraphics[scale=0.58]{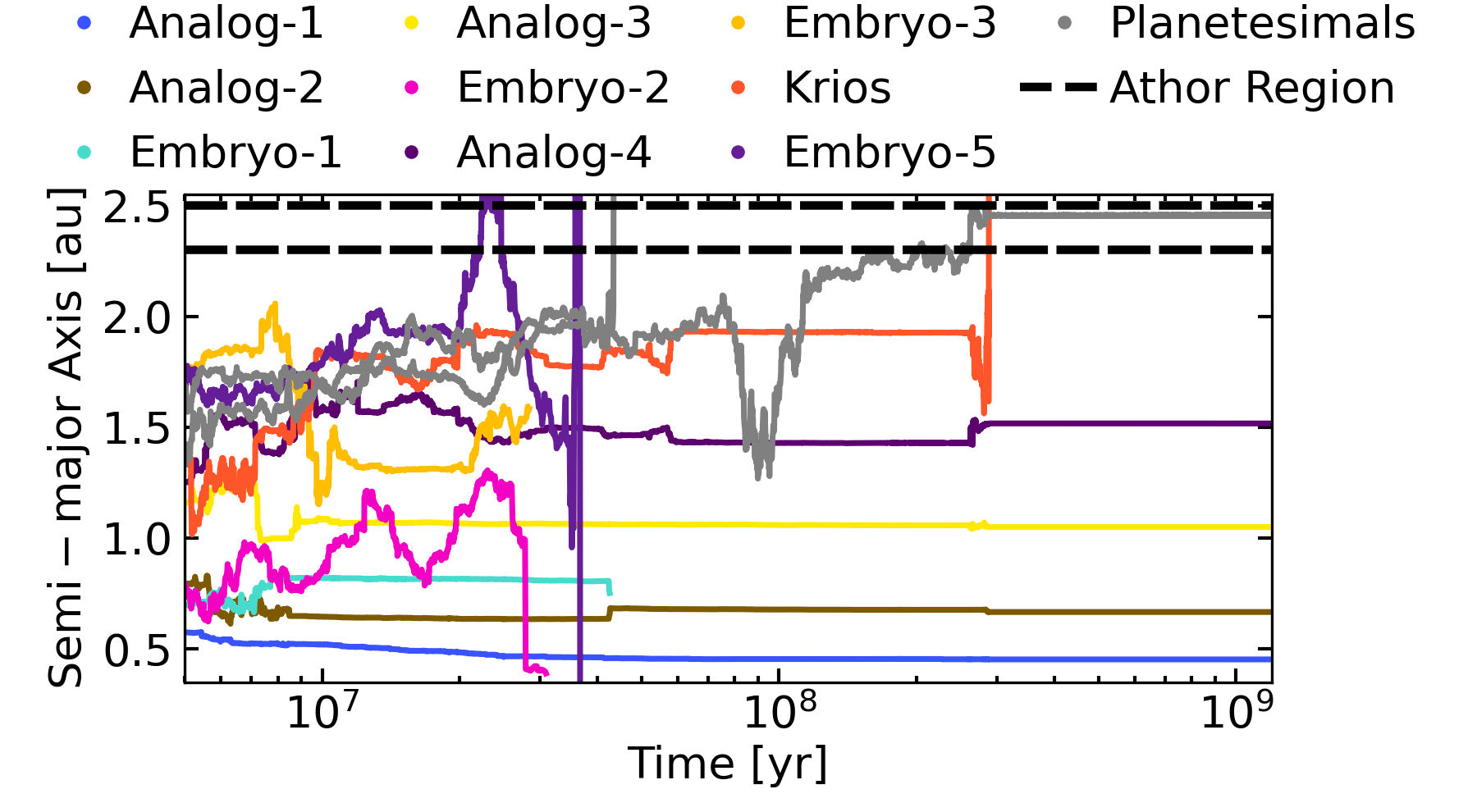}
\includegraphics[trim={0 0 0cm 1.5cm},clip,scale=0.58]{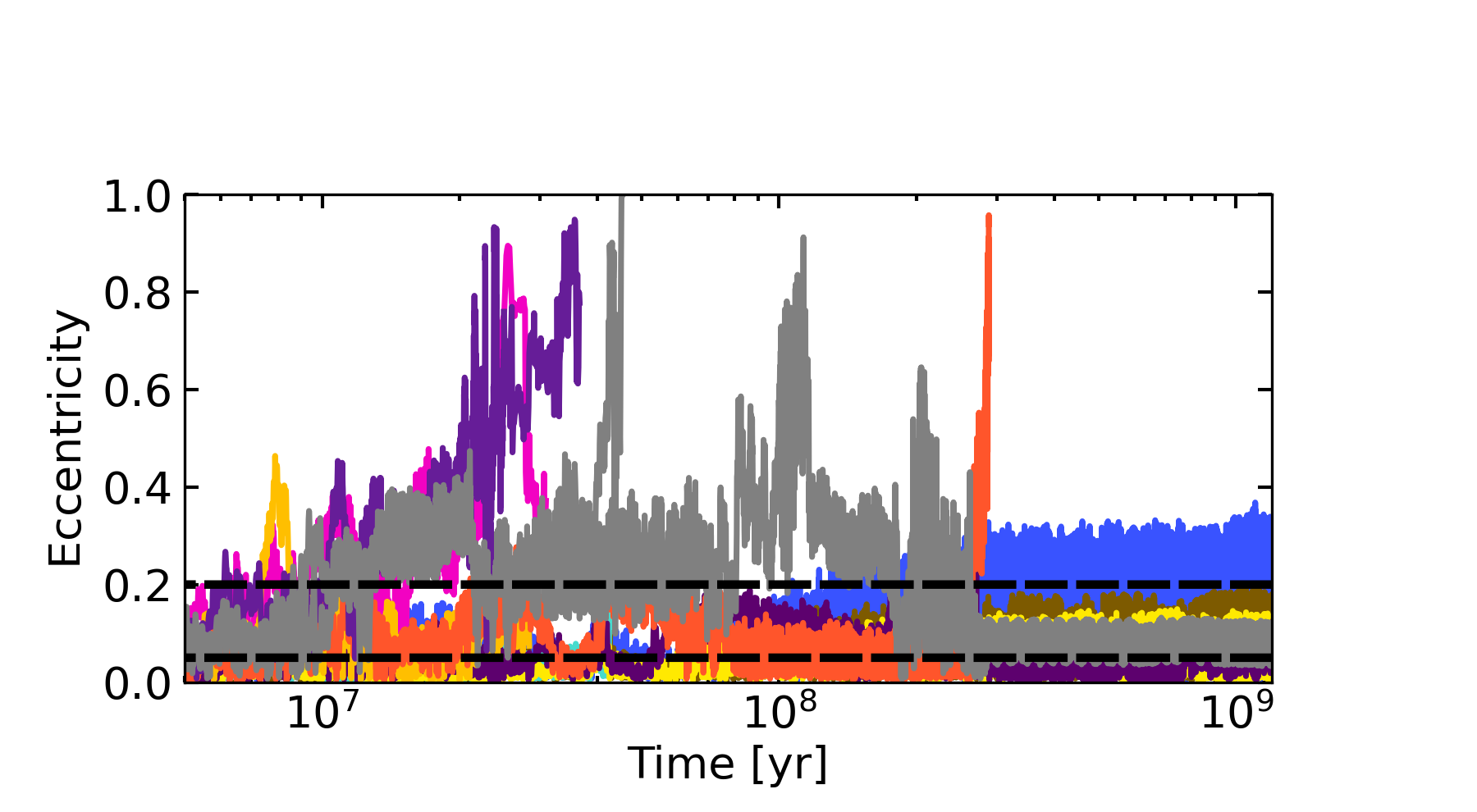}
\includegraphics[scale=0.58,trim={0 0 0cm 1.5cm},clip]{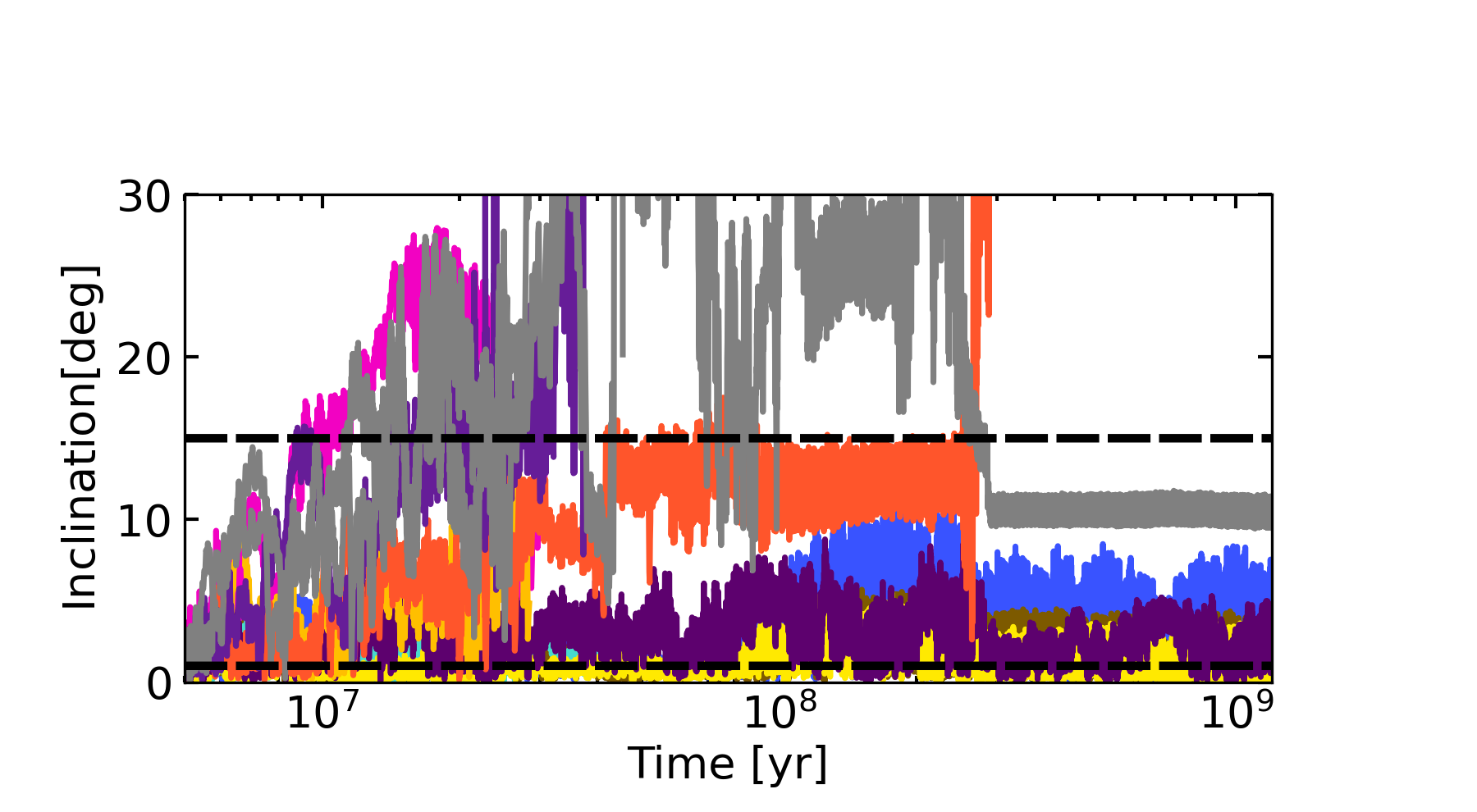}
\caption{Implantation of a planetesimal from the terrestrial region into the asteroid belt during the final accretion phase of terrestrial planets in our Athor-002 simulation. The planetesimal is implanted in the inner main belt, in the Athor-region. The x-axes shows time and the y-axes show, from top-to-bottom, semi-major axis, orbital eccentricity, and inclination. Dashed lines  delimit the Athor region defined as $2.3<a<2.5$~au, $e<0.15$, and $i<15$~deg. Like figure \ref{fig:analog1}, this simulation corresponds to another example where the instability takes place at the time of the disk dispersal, i.e., $t_{\rm inst}=$5~Myr after the solar system formation. At 5~Myr, the grey-line planetesimal is at about 1.6~au and it finally enters the asteroid belt at $\sim$100 Myr ($a\gtrsim1.8$~au). Its effective implantation  occurs at $\sim$290~Myr, when Krios (orange) is ejected from the system.}
    \label{fig:analog2}
\end{figure}

We performed a second set of simulations where we selected another simulation with $t_{\rm inst}=$5~Myr that produced an Athor analog and cloned it 120 times. We refer to this simulation as ``Athor-2''. We cloned this system at the 25 Myr point of our original simulation, with the goal of improving our statistics and motivated by the fact that the system at 25 Myr looks like a good candidate for solar system analog. Out of our 120 simulations, one implanted  a planetesimal in the Athor region. The dynamical evolution of this system is shown in Figure \ref{fig:analog2}. This clone simulation produced four terrestrial planets (Analogs 1 to 4). In  Figure \ref{fig:analog2}, the implanted planetesimal finally entered the asteroid belt at about 105 Myr but it did not attain its final, fully implanted semi-major axis and eccentricity until  $\sim$290~Myr. It started initially at about 1.4~au  at 5 Myr, reached the asteroid belt region (a$>$1.8 au)  for the first time  between 15 and 20 Myr,  left the main belt region again between 65 and 100 Myr, and finally re-entered the main belt at about 105 Myr. This late implantation  makes this object consistent with Athor~\citep{avdellidouetal24}.

A few other clone simulations of Athor-2 implanted planetesimals in the belt with slightly higher orbital eccentricities and inclinations than our favourite case (and Athor). We do not consider these cases when computing implantation probabilities. The implantation probability of our best Athor-analogue by itself, accounting for all  120 clone simulations (1 implanted planetesimal / (120 clones$\times$500 planetesimals), is about $\approx1.7\times10^{-5}$ for an instability with $t_{\rm inst}=$5~Myr. Accounting for these additional slightly less appreciable implantation cases the overall implantation efficiency could be up to 2 to 3 times higher.

%A small level of damping of eccentricity and inclination after the ejection of ``Embryo-2'' is observed for Mars and the implanted planetesimal due to the dynamical effects of leftover planetesimals in the system. These planetesimals are not shown in the figure for presentation purposes.

Our  Athor-1 and Athor-2  simulations produced our best examples of solar system analogues containing Athor-like objects.  Both our  Athor-1  and Athor-2   simulations match our defined solar system analogue constraints. We have verified that the two Mars-analogues of Figures \ref{fig:analog1} and  \ref{fig:analog2}  grew relatively fast and were almost fully formed (reaching $>$ 90-95\% of their final masses) during the gas disk phase. This makes them broadly consistent with estimated Mars'  accretion timescale \citep[e.g.][]{dauphaspourmand11}.

Given the combined results of our instantaneous and interpolated instability (see previous section) simulations, we conclude that, if the instability occurred within 15 Myr of solar system formation, as in our Athor-1 and Athor-2 simulations, the probability of implanting an Athor-like planetesimal into the main belt  after $>$60 Myr was only $\sim$2-3 times lower than the probability of implantation before $<$60 Myr.

%Our Athor-1 and Athor-2 simulations represent specific cases with early instability times, i.e.,  $t_{\rm inst}=$5~Myr.

Until now we have dedicated this section to analyse and discuss the results  of simulations  with early instability times, i.e.,  $t_{\rm inst}=$5~Myr and 15 Myr. Now we analyze the implantation probability in simulations with late instabilities, $t_{\rm inst}=$55 and 105~Myr. 

Out of our 200 simulations with $t_{\rm inst}=$55~Myr, 4 planetesimals were implanted in the Athor region, resulting in an implantation efficiency of $\approx5\times10^{-5}$. However, all these four planetesimals first entered the asteroid belt (a$>$1.8~au) between 15 and 40 Myr after CAIs, with an average entrance time of $\sim$25 Myr, making them inconsistent with Athor~\citep{trieloffetal22}.

Out of our 200 simulations with $t_{\rm inst}=$105~Myr, a total of 13 planetesimals were implanted in the Athor region of the belt.  Out of these 13 planetesimals 4 were implanted in the asteroid belt after 60 Myr, with an average time of entrance in the belt of $\sim$ 70 Myr after CAIs. This yields an implantation probability of Athor-analogues of about  $\approx5\times10^{-5}$, which is marginally consistent with the estimate derived from our interpolated simulations (see previous section). Confirming our results of the previous section, these results also suggest that implanting objects in the Athor region at $\sim$100 Myr with $t_{\rm inst}=$105~Myr has a probability of the order of $\lesssim1.2\times10^{-5}$.

%(We will combine this probability with others derived in our simulations below.)

\subsubsection{The Implantation mechanism}

\begin{figure}[h]
\centering
\hspace{-1.1cm}\includegraphics[scale=0.33]{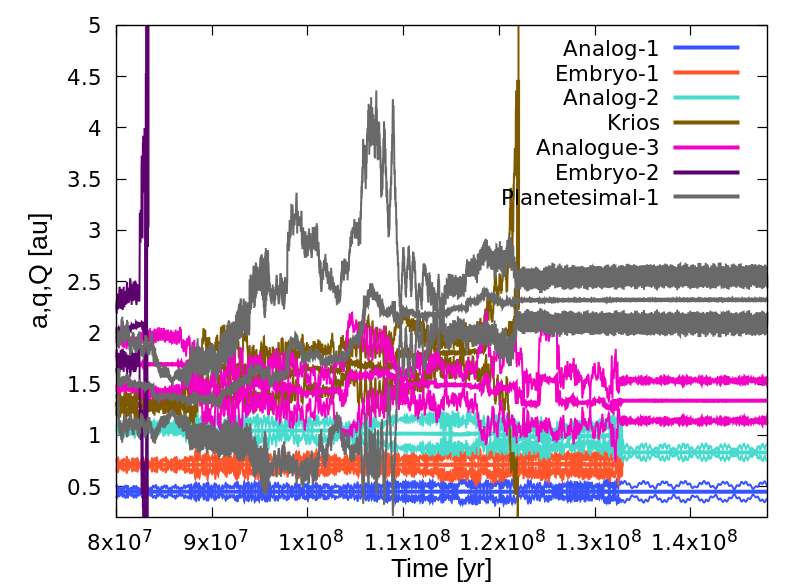}
\caption{Evolution of semi-major axis, pericenter, and apocenter of all planetary bodies shown in Figure \ref{fig:analog1}, using the same color-coding in both figures. The implantation of the planetesimal in the belt (grey-lines object) at about 120~Myr occurs due to its gravitational interaction with Krios and Mars. The high eccentricity and inclination phases of Krios before its ejection induce secular effects on the planetesimal that reduce its orbital inclination and eccentricity, ultimately implanting it in the belt around Athor's modern orbit.}
    \label{fig:orbit_analog1}
\end{figure}

Athor analogs implanted after the giant planet instability all have similar origins.  This is true regardless of whether they are implanted as late as those in Figures \ref{fig:analog1} and \ref{fig:analog2} or earlier (see examples in the Appendix). They are typically planetesimals scattered from $\approx$1-1.7~au into the Mars-region that undergo multiple scattering events by a Mars-analogue and a leftover unstable planetary embryo on an asteroid-belt-crossing orbit. We refer to this embryo as {\it Krios}, the brother of {\it Theia}, the name commonly given to the Moon-forming impactor. These objects are shown in Figures \ref{fig:analog1}, \ref{fig:analog2}, and \ref{fig:orbit_analog1}.
We also found implantation examples in our sample of simulations where Theia assumes the role of ``Krios''. In these specific cases, Theia, rather than being ejected from the solar system like Krios, collides with the Earth-analogue while also promoting Athor's implantation.

Figure \ref{fig:orbit_analog1}  shows the evolution of the semi-major axis, pericenter, and apocenter of all objects shown in Figure \ref{fig:analog1}. Although orbital overlap between the planetesimal and the asteroid belt exists for $<$60 Myr, this planetesimal may have avoided break up to 80 Myr or so, when it may have finally fragmented near its pericenter. Note that at 85~Myr it is outside the belt ($a<Q<1.8$~au). In this case, as all  fragments would be in terrestrial planets crossing orbits, all but one (the lucky Athor) would have been lost.

In our simulations, Athor-like objects were typically implanted into the asteroid belt after experiencing a secular torque from Krios (or Theia), typically during Krios' (Theia's) final high eccentricity/inclination phase (see Figure \ref{fig:orbit_analog1} and an additional example in the Appendix).

Our simulations show that planetary embryos like Krios play a fundamental role in implanting Athor-like objects in the asteroid belt. The simulations of \cite{avdellidouetal24} did not capture the effect of this type of object during the final phase of accretion of terrestrial planets. Their simulations did not implant planetesimals in the Athor region for early instability times due to assumptions in their design and very limited number of simulations. \cite{avdellidouetal24} tested four different implantation scenarios: i) implantation after terrestrial planet formation, where the terrestrial planets and giant planets are assumed to be fully formed and in their current orbits; ii) implantation during terrestrial planet formation, using a single simulation outcome from \cite{nesvornyetal2021}; iii) implantation due to the gravitational effects of a leftover planetary embryo envisioned to be Theia; and iv) implantation during the giant planet instability. \cite{avdellidouetal24}  ruled out scenarios i), ii), and iii) and concluded that only iv) is plausible. Our results support their conclusions that scenario i) can not implant Athor like objects. However, we show here that scenario ii) and iii) are also consistent with Athor. Their simulation of scenario ii) probably failed to implant planetesimals in the Athor region because they only tested a single accretion history of the terrestrial planets. Their scenario iii) failed to implant planetesimals in the Athor region because they assumed an ad-hoc initial orbit for the putative leftover embryo. 

%The idealized simulations of \cite{avdellidouetal23} failed to capture the effect of this type of object during the final phase of accretion of terrestrial planets. Their simulations failed to implant planetesimals in the Athor region as consequence of a flaw in design.

%The Mars-analogue in this case is about 1.75 Mars-mass and the Earth-analogue is about 0.7$M_{\oplus}$. Krios, represented by Embryo-2, undergoes a close-encounter with a planetesimal (grey lines) at about 90~Myr, implanting the planetesimal in the Athor-region. Krios is subsequently ejected from the solar system due to a close-encounter with Jupiter. In this particular simulation the instability occurred at the time of the disk dispersal (time zero of the simulation) but simulations with  instabilities at $t_{\rm inst}=$10~Myr produced similar results (see Figure \ref{fig:Athor-like2}). 

\subsubsection{Which instability timing is more likely to implant Athor?}

Our simulations with an instability occurring within 15 Myr after CAIs formation yield a probability of implanting Athor after 60 Myr of at least $\approx2\times10^{-5}$.
Simulations with an instability occurring at 105 Myr after CAIs formation produce implantation probability of 
$\approx5\times10^{-5}$. This shows that the probability of implanting Athor in the early instability scenario \citep{clementetal18,nesvornyetal2021,liuetal22} is at most a factor of $\sim$2.5 lower than that of implanting it in the late instability. We argue that a simple difference of about a factor of two is insufficient to rule out one dynamical scenario over the other.  Indeed, simulations of the giant planet instability itself have a probability of matching the present-day giant planets' orbits that is significantly lower than 50\%~\citep[e.g.][]{nesvornymorbidelli12,clementetal21}.  If one were to make judgments on past events in Solar System history based on factors of two, the instability itself would be thrown out.

\begin{figure}
\hspace{-1cm}
\includegraphics[scale=0.4]{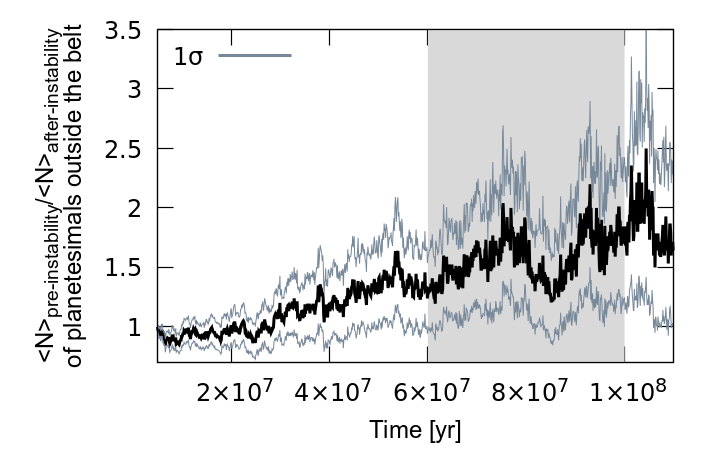}
\caption{Temporal evolution of the ratio of the average number of planetesimals outside the main asteroid belt in simulations where the instability takes place at 5 Myr (giant planets are in their current orbits) after solar system formation and simulations where the giant planets are kept in pre-instability orbits. Planetesimals outside the belt are defined as those in the terrestrial region ($1<a<1.8$~au) and those in the belt region but with very high orbital eccentricities and inclinations ($1.8<a<3.5$~au, and $e>0.4$, and $i>35$~deg). We have tested several different cutoffs for the definition of planetesimals outside the main belt and the results were similar to the presented case. For instance, accounting only for planetesimals in the terrestrial region ($1<a<1.8$~au) leads to broadly similar results. The number of planetesimals outside the belt for each giant planet configuration is averaged over 50 simulations modelling the accretion of terrestrial planets. The grey-lines show 1-$\sigma$ standard deviation.}
    \label{fig:planetesimal_ratio}
\end{figure}

Our implantation probabilities -- and the factor of $\sim$2 difference for Athor' s implantation in a late vs early instability --  can be more easily understood if we recall that, to first order, derived implantation efficiencies are a function of the total number of planetesimals available for implantation at different times. We have computed the ratio of the number of planetesimals existing outside the belt in simulations with an instability occurring at 5 Myr and those where the giant planets are kept in pre-instability orbits. Our results are presented in Figure \ref{fig:planetesimal_ratio}.

%Our conclusion is further supported by the fact that this inferred factor of two difference in implantation efficiencies is in agreement with the evolution of the   By counting the number of existing planetesimals outside the belt as the system evolves, we can obtain an order of magnitude estimate of which scenario is more likely to implant Athor after 60 Myr.

%e recall the reader that Athor's parent body is likely to have fragmented after 60 Myr outside the belt~\citep{avdellidouetal22,avdellidouetal23} and the instability time is constrained to be 

Figure \ref{fig:planetesimal_ratio} shows that the inferred factor of two difference in implantation efficiencies between our early and late instability simulations is largely in agreement with the evolution of the ratio of the number of planetesimals outside the asteroid belt in these simulations. As one can see, simulations with giant planets in  pre-instability orbits tend to have a factor of $\sim$1.4 to $\sim$2 more planetesimals available outside the belt between 60 and 100 Myr than those where they have attain their modern orbits at 5 Myr after CAIs.  Once Jupiter and Saturn are placed on their present-day orbits, they slightly accelerate the process of accretion and depletion in the terrestrial region.  Thus, the number of planetesimals available to be implanted on Athor-like orbits is reduced quicker than when the giant planets are on their pre-instability orbits. Therefore, Figure \ref{fig:planetesimal_ratio} supports our findings and suggests that the inferred mere factor of $\sim$2 difference in the probabilities of implanting Athor with an instability at $<15$~Myr compared to an instability at $>$60~Myr is reasonable and robust. From our results, we can not firmly exclude the possibility that these probabilities are virtually the same, in particular for implantation times $\gg70$Myr.

We conclude this section by re-emphasizing that the link between Athor and EL meteorites suggested by \cite{avdellidouetal22,avdellidouetal24} is not sufficient to constrain the timing of the giant planet dynamical instability.

%Only our simulations with  $r^0$ and  $r^{-1}$ were able to implant Athor-like objects in the belt for  $t_{\rm inst}=$ 5~Myr and 10~Myr. Our simulations with radially steep rings were less efficient in implanting Athor-like objects in the belt at the early times because embryos growing beyond 1~au are still too small to significantly scatter planetesimals into the belt \citep{raymondizidoro17b,izidoroetal22}. We will now look for  Athor-like planetesimals in our ``instantaneous''  instability simulations modelling the late stage of accretion of terrestrial planets.

%our ``instantaneous''  instability simulations to demonstrate that Athor-like planetesimals can be also implanted after the instability have occurred, during the accretion and chaotic evolution of the growing terrestrial planets.
%In this subsequent analysis,  we  select only systems produced in simulations invoking the instantaneous instability approach with  $t_{\rm inst}=$ 0 and 10~Myr. We are interested in simulations that simultaneously implant Athor-like objects and produce reasonable solar system analogues. 

\section{Conclusion}\label{sec:conclusion}

In this paper we studied the implantation of the asteroid Athor from the terrestrial region into the asteroid belt. We were motivated by the recent results of \cite{avdellidouetal24} suggesting that Athor can not be implanted in the belt if the giant planet dynamical instability occurred before 60 Myr after the formation of the solar system. To revisit this problem, we  performed and analyzed a suite of numerical simulations modelling the accretion of terrestrial planets from narrow ring of planetesimals \citep{izidoroetal24}, with a focus on the implantation of planetesimals from the terrestrial region into the asteroid belt.
 
We modelled the giant planet dynamical instability by assuming that the giant planets formed in a more compact, circular, and coplanar configuration, before evolving to their current dynamical state. The timing of the instability was treated as a free parameter in this work. We performed simulations assuming that the giant planet instability occurred at 5, 10, 15, 55, and 105 Myr after the formation of the solar system. 

Our simulations showed that Athor-like objects can be implanted in the asteroid belt up to $\gtrsim$100 Myr after the solar system dynamical instability occurred. We compared the probability of implanting Athor after 60 Myr in scenarios where the instability occurs ``early'', within 15 Myr, and relatively ``late'', at $\sim$100 Myr after the solar system formation. Our results show that the probability of implanting Athor-like objects when the instability occurs at $\sim$ 100 Myr is at most a factor of 2 higher than that from the scenario where the instability occurs within 15 Myr after the formation of the solar system.

We conclude this work by stating that Athor's origin alone can not be used to constrain the time of the solar system dynamical instability.  A dynamical instability occurring at $\lesssim$15~Myr after the formation of the solar system remains a plausible scenario to explain both the implantation of the asteroid Athor, as well as the total mass carried by S-complex  asteroids \citep[taxonomic class, i.e.,][]{demeocarry14} in the belt \cite[see][]{izidoroetal24}. A very early instability could have been triggered by dynamical interactions between the giant planets themselves~\citep{ribeiroetal20} or by dispersal of the gaseous disk~\citep{liuetal22}.  Such a scenario has already been shown to be compatible with thermochronological measurements of asteroidal meteorites~\citep{edwards23}, and the dynamical influence of an early instability could explain a number of features of Earth, the terrestrial planet system, and asteroid belt~\citep{clementetal18,clementetal2019,nesvornyetal2021,clement23,joiret23}.

\vspace{1cm}
%\begin{acknowledgements}
 {\bf Acknowledgements:} The work of R.D. was supported by the NASA Emerging Worlds program, grant 80NSSC21K0387. MSC is supported by NASA Emerging Worlds grant 80NSSC23K0868 and NASA’s CHAMPs team, supported by NASA under Grant No. 80NSSC21K0905 issued through the Interdisciplinary Consortia for Astrobiology Research (ICAR) program. SNR thanks the CNRS's PNP and MITI/80PRIME programs for support. A.~I thanks Debjeet Pathak for several discussions and  proofreading the paper. 

\bibliography{mybib}{}
\bibliographystyle{aasjournal}

\clearpage
\appendix
\section{Additional Examples of implantation of planetesimals in the Athor-region}
\begin{figure}[h]
\centering
\includegraphics[scale=0.68]{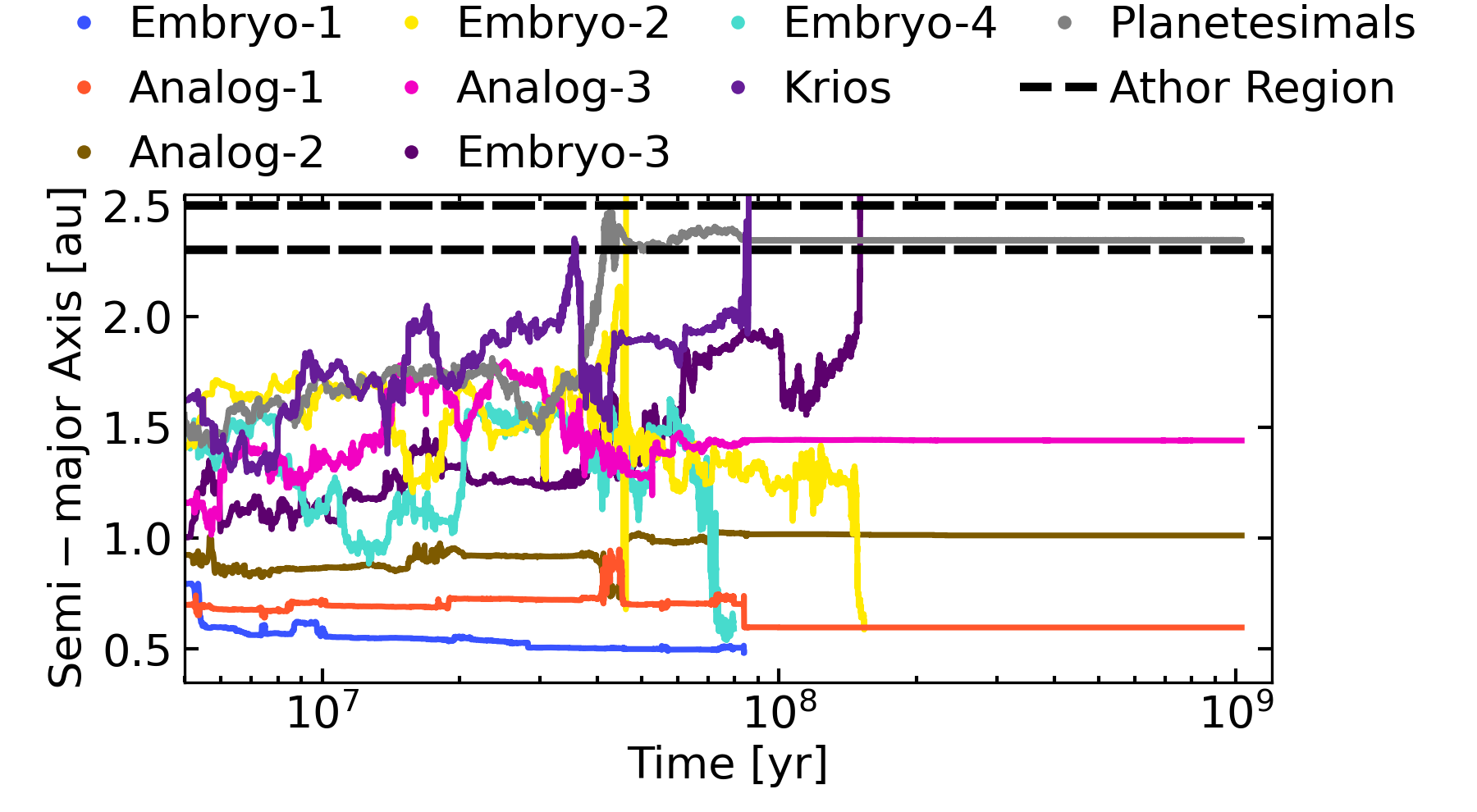}
\includegraphics[trim={0 0 0cm 1.5cm},clip,scale=0.68]{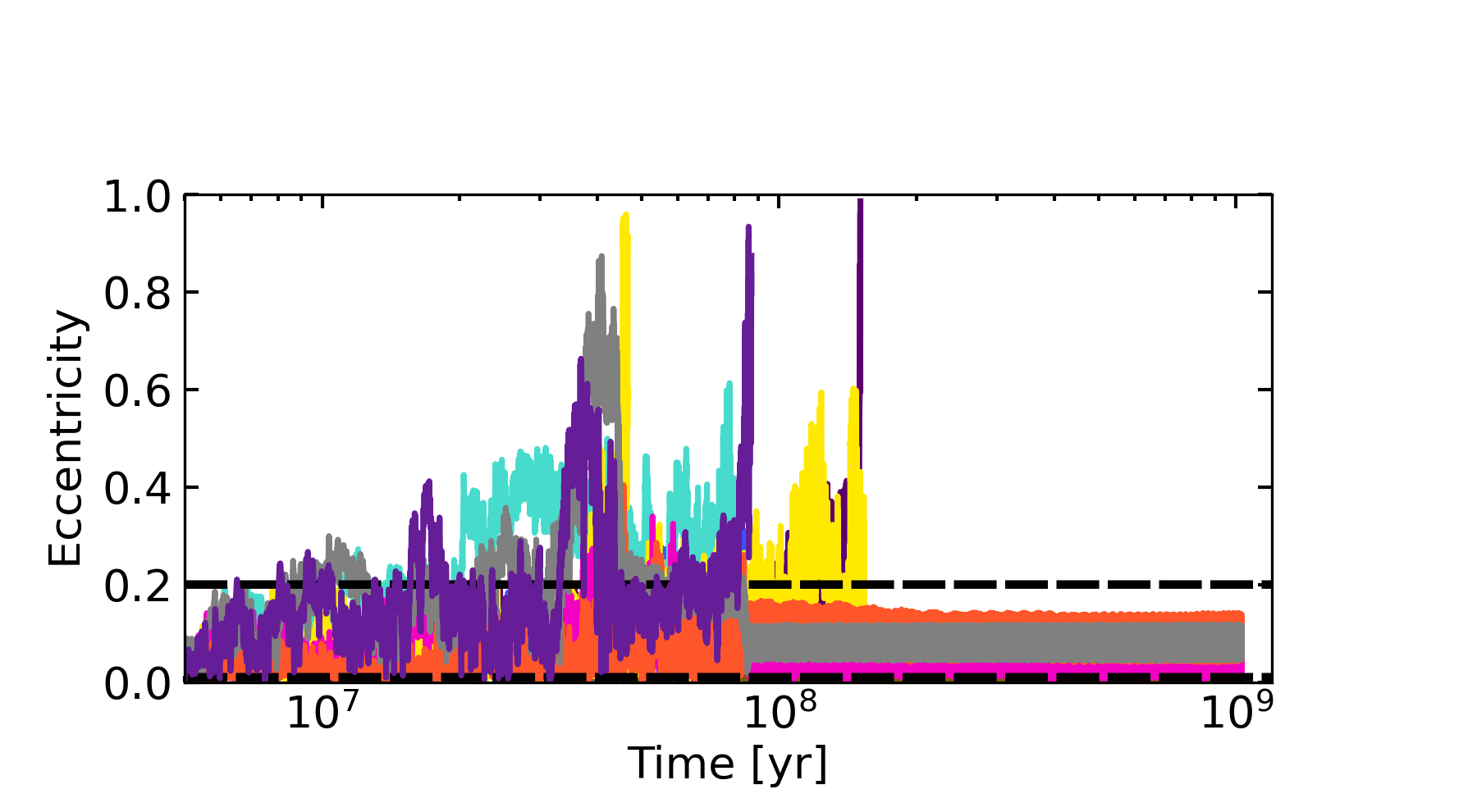}
\includegraphics[scale=0.68,trim={0 0 0cm 1.5cm},clip]{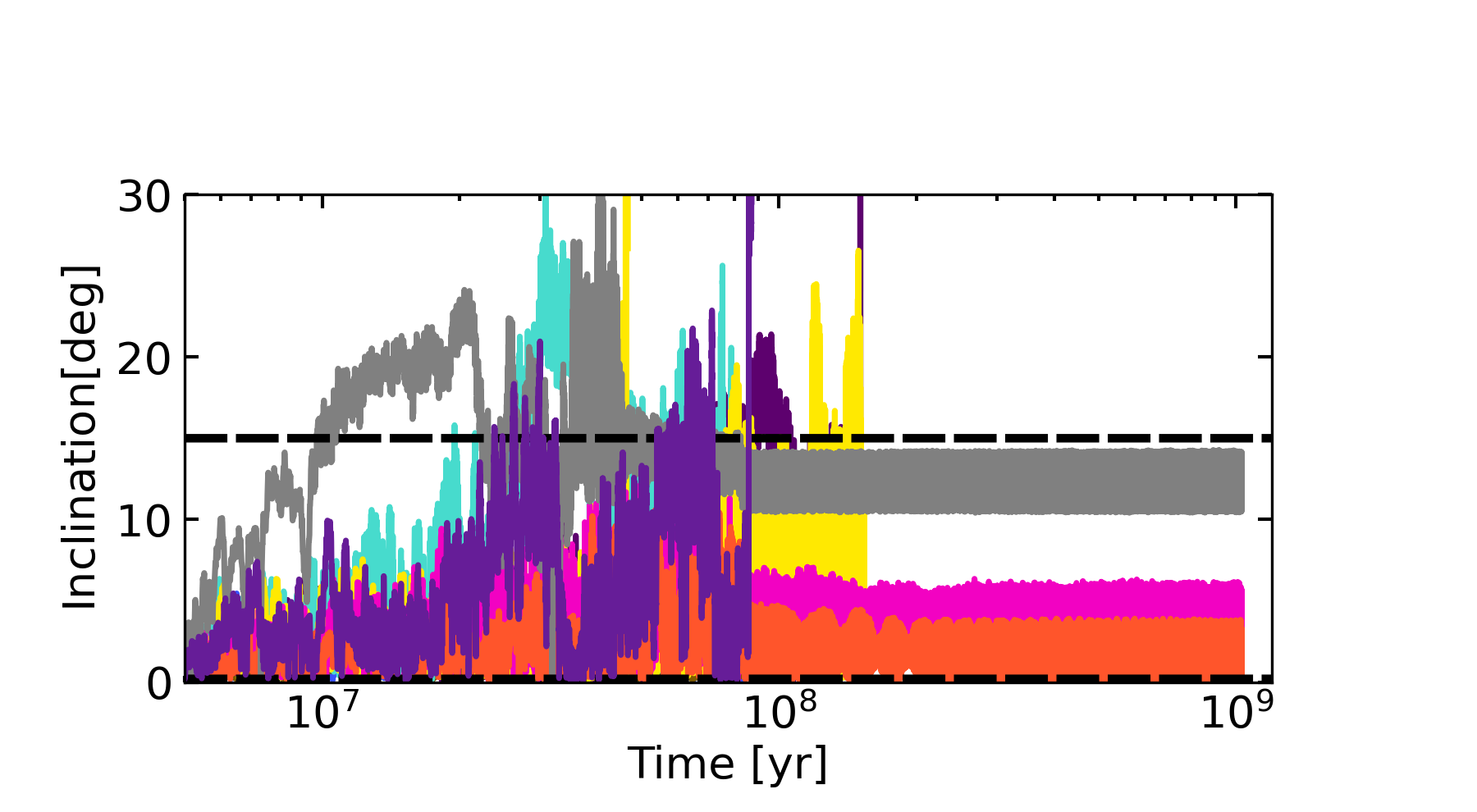}
\caption{Representative example of implantation of a  planetesimal that arrives relatively early in the belt, at about $\sim$40~Myr.
The early arrival of this object in the asteroid belt makes it most likely inconsistent with Athor. As in Figure \ref{fig:analog1}, the x-axes shows time and the y-axes shows, from top-to-bottom, semi-major axis, orbital eccentricity, and inclination. Dashed lines  delimit the Athor region defined as $2.3<a<2.5$~au, $e<0.15$, and $i<15$~deg. This simulation corresponds to an example where the instability also takes place at the time of the disk dispersal, $t_{\rm inst}=$5~Myr relative to the solar system formation time. The final implantation of this planetesimal happens at about 85~Myr when Krios is ejected from the system.}
    \label{fig:Athor-early-1}
\end{figure}

\begin{figure}[h]
\centering
\includegraphics[scale=0.68]{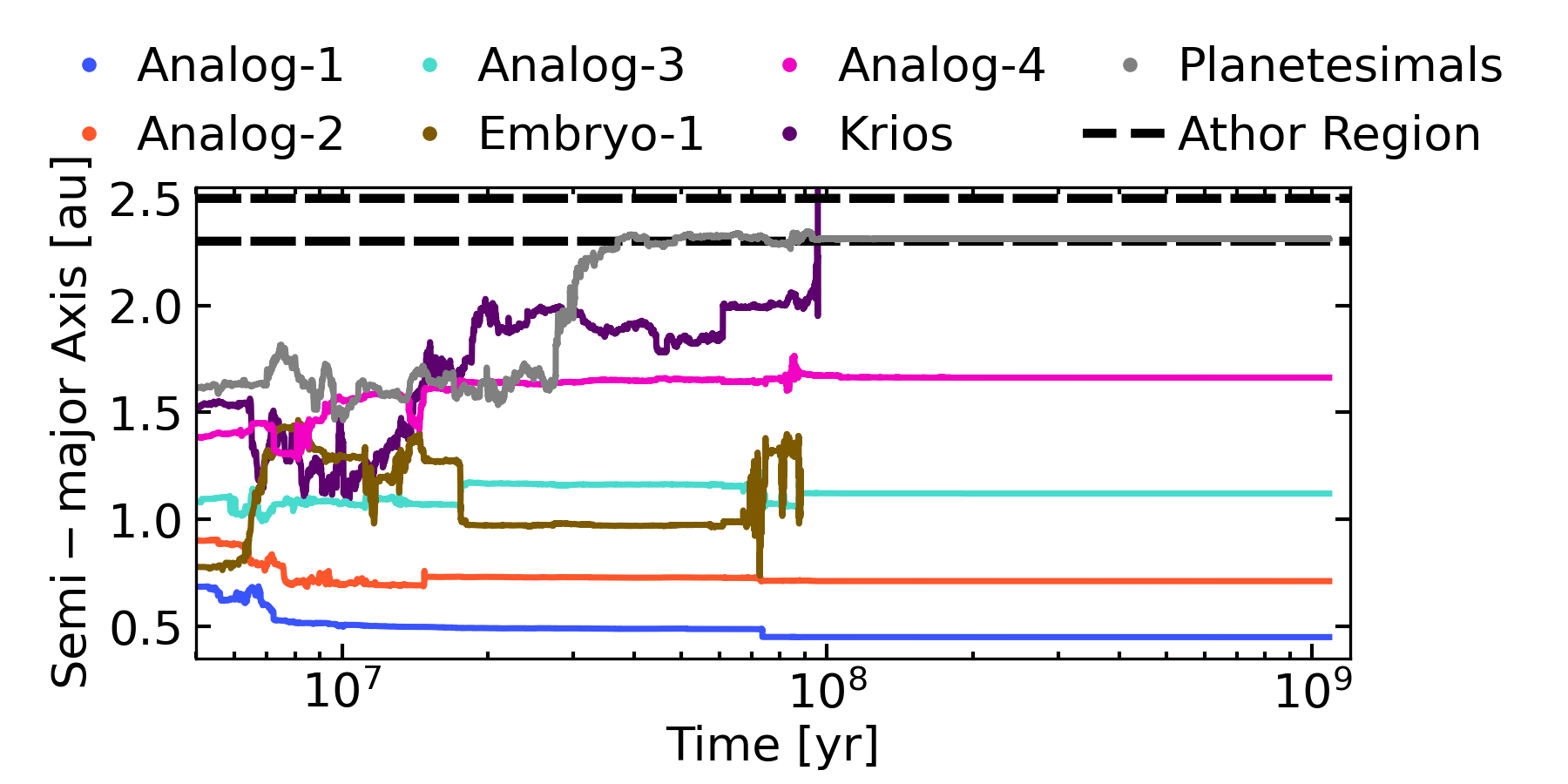}
\includegraphics[trim={0 0 0cm 1.5cm},clip,scale=0.68]{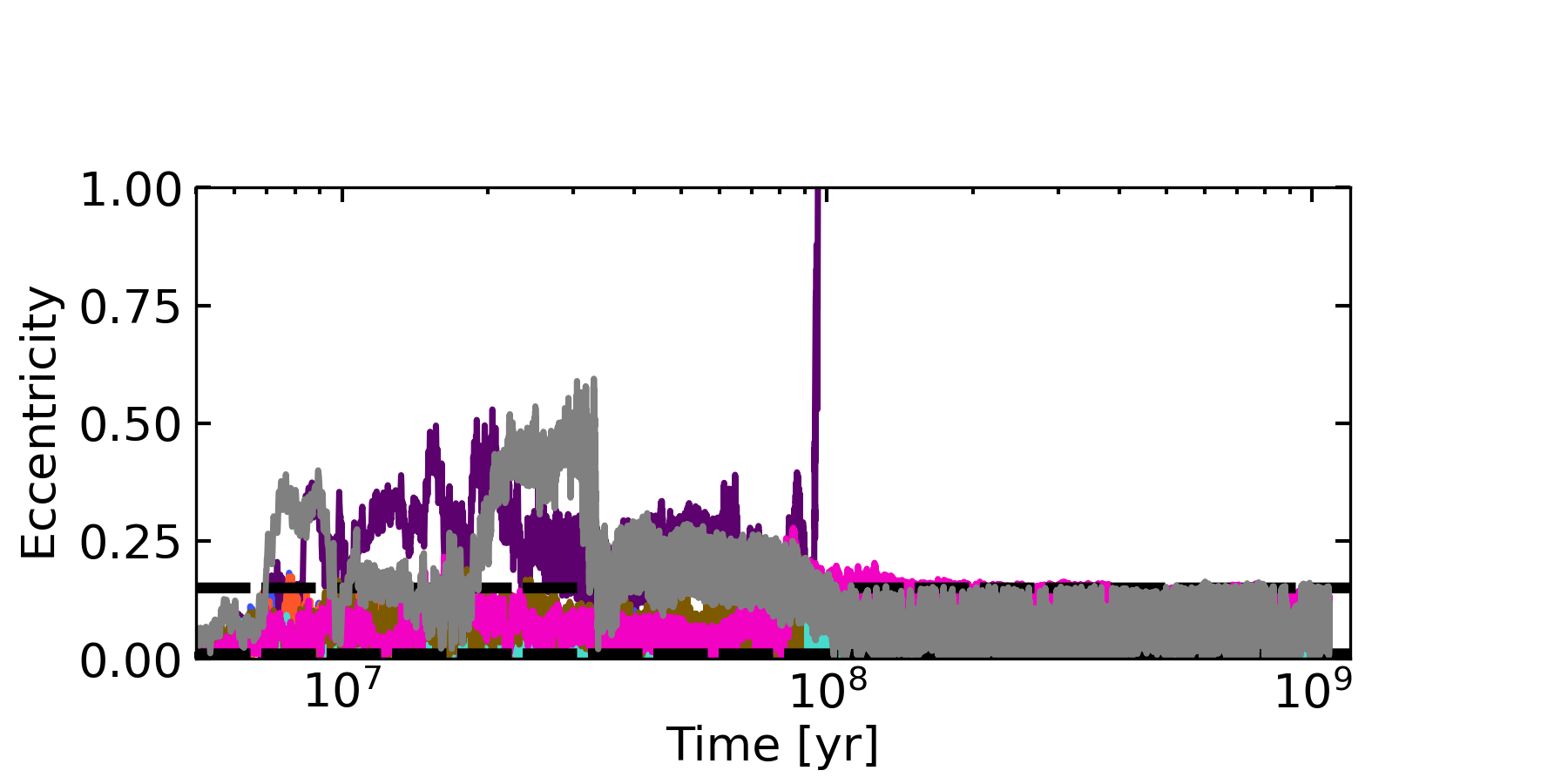}
\includegraphics[scale=0.68,trim={0 0 0cm 1.5cm},clip]{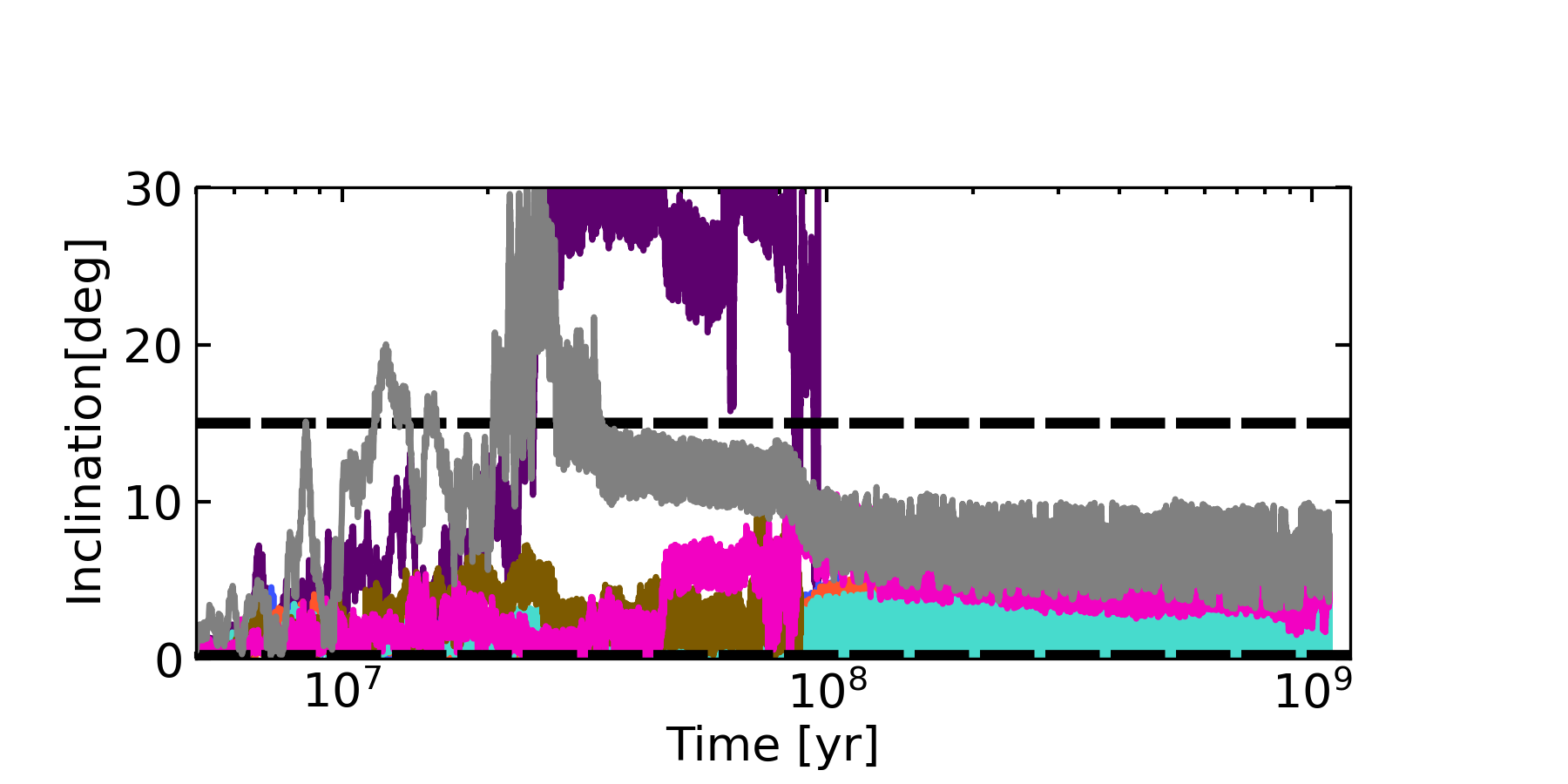}
\caption{Representative example of implantation of a  planetesimal that arrives relatively early in the belt, at about $\sim$40~Myr.
The early arrival of this object in the asteroid belt makes it most likely inconsistent with Athor. As in Figure \ref{fig:analog1}, the x-axes shows time and the y-axes shows, from top-to-bottom, semi-major axis, orbital eccentricity, and inclination. Dashed lines  delimit the Athor region defined as $2.3<a<2.5$~au, $e<0.15$, and $i<15$~deg. This simulation corresponds to an example where the instability also takes place at the time of the disk dispersal, $t_{\rm inst}=$5~Myr relative to the solar system formation time. The final implantation of this planetesimal happens at about 85~Myr when Krios is ejected from the system.}
    \label{fig:Athor-early-2}
\end{figure}

Examples presented in this section show implantation of planetesimals in the Athor region in simulations that also produced good solar system analogues. Although in both Figures \ref{fig:Athor-early-1} and \ref{fig:Athor-early-2} planetesimals  were only effectively implanted in the belt at about 100~Myr, both implanted objects arrived in the belt region too early (before 60 Myr after CAIs formation), making them inconsistent with Athor.

\begin{figure}[h]
\centering
\includegraphics[scale=0.35]{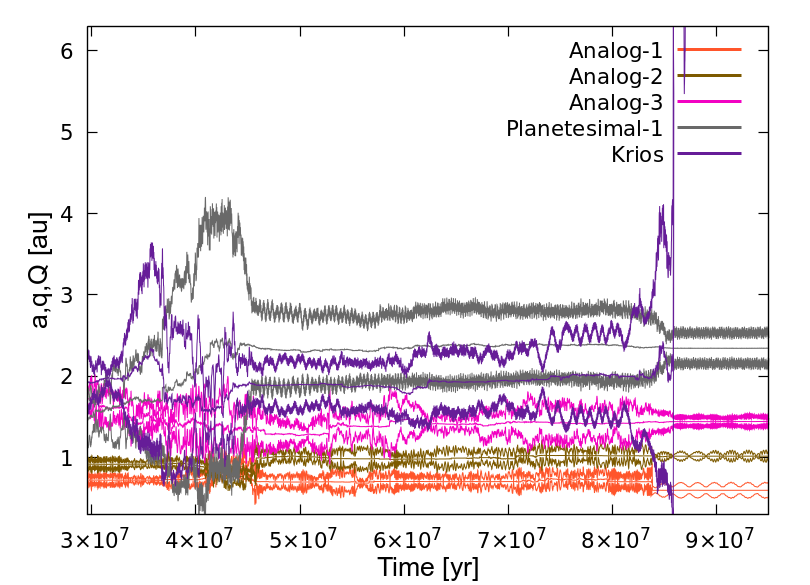}
\caption{Evolution of semi-major axis, pericenter, and apocenter of all planetary bodies shown in Figure \ref{fig:Athor-early-1}, using the same color-coding for both figures. The implantation of the planetesimal in the belt (grey-lines object) at about 85~Myr occurs due to its gravitational interaction with Krios and Mars, but note that it enters the asteroid belt region ($a\gtrsim1.8$~au) between 30 and 40 Myr. As in Figure \ref{fig:orbit_analog1}, the high eccentricity and inclination phases of Krios before its ejection induce secular effects on the planetesimal, reducing its orbital inclination and eccentricity, and ultimately implanting it in the belt and Athor's region.}
    \label{fig:orbit-early-1}
\end{figure}

\end{document}